\documentclass[%
 reprint,
 amsmath,amssymb,
 aps,
]{revtex4-1}

\usepackage{graphicx}
\usepackage{dcolumn}
\usepackage{bm}
\usepackage{comment}

\begin{document}

\preprint{APS/123-QED}

\title{Heavy baryons in holographic QCD with higher dimensional degrees of freedom}

\author{Daisuke Fujii}
 \email{daisuke@rcnp.osaka-u.ac.jp}
\author{Atsushi Hosaka}%
 \altaffiliation[Also at ]{Advanced Science Research Center, Japan Atomic Energy Agency, Tokai, Ibaraki 319-1195 Japan.}
  \email{hosaka@rcnp.osaka-u.ac.jp}
\affiliation{%
 Research Center for Nuclear Physics, Osaka University, Ibaraki 567-0048, Japan.\\
}%

\date{\today}

\begin{abstract}
We try to introduce heavy flavors to Sakai-Sugimoto model by regarding higher dimensional components of gauge fields as heavy mesons. 
Using the Forg\'acs and Manton approach, we obtain a theory composed of heavy mesons and the instanton of light flavors. 
Applying the collective coordinate quantization method, we derived a mass formula of heavy baryons. 
In the leading order of $1/m_{H}$ expansion ($m_{H}$ the mass of a heavy quark), we find singlet and doublet states in the heavy quark symmetry (HQS). 
Also, we obtain the degenerate Roper like and odd parity excitations. 
By virtue of heavy meson degrees of freedom, our mass formula reproduces the mass ordering of $\Sigma^*_c$ and $\Lambda_c^*$ correctly.

\end{abstract}

\maketitle

\section{\label{Introduction}Introduction}

In the past decades, hadron physics has experienced many new findings of exotic phenomena mostly containing charm and bottom (heavy) quarks 
that are not easily explained by conventional approaches, 
which is telling that we need new ideas~\cite{Sakai2005,2Sakai2005}, and references there in~\cite{Yamaguchi:2019vea,Hosaka:2016pey,Chen:2016qju,Tanabashi:2018oca}.  
Due to non-perturbative nature of  the strong interaction dynamics, there still remains a problem of missing link from quarks to hadrons. 
Holographic view of QCD has attracted much attention as one of guiding principle to fill that link. 
Among several alternatives, Sakai and Sugimoto proposed a D4-D8 brane construction that has lead to low energy effective actions for hadrons, 
where the model reproduces important features of spontaneous breaking of chiral symmetry including hadron resonances 
such as $\rho$ and $a_1$ mesons with a few parameters~\cite{Sakai2005,2Sakai2005}. 

The Sakai-Sugimoto model is a nine-dimensional gauge theory of flavor SU(2) on the D8-brane in the D4-brane background. 
Our model is an SU(2+1) extension of the Sakai-Sugimoto model. 
The gauge fields are denoted as $A^a_M$ 
where  $a=1, ... , 8$ are for the flavor index and $M = 0, ... , 3, z, 6, ... , 9$  for gauge field components. 
In comparison with actual QCD, Sakai and Sugimoto have utilized five components, $A_{M = 0, ..., 3, z}^a$, while the other four 
were ignored when they derived a five-dimensional gauge theory~\cite{Sakai2005}. 
Then the fifth-dimensional degrees of freedom play a role of generating 
various hadron resonances of light flavors of $u, d$ quarks in the four-dimensional space-time. 

In this paper, using the extra-dimensional degrees of freedom, we attempt to construct a model with heavy flavors for the study of heavy baryons. 
The gauge field living in the extra higher dimensional space-time is transformed into heavy mesons by the method of 
dimensional reduction by Forg\'acs and Manton~\cite{Manton1980,Manton1979}. 
The method leads to the field components that correspond to heavy mesons and light flavor instantons. 
Supplemented by a mass term, we propose a set of actions for heavy hadrons. 

Once we establish the model in five-dimensional space-time, we apply the standard method to the quantization of 
instantons for heavy baryons,  from which we compute various physical quantities. 
We estimate masses of heavy baryons that are compared with existing experimental data. 
We would then like to study whether such a construction 
provides a reasonable description for not only conventional but also 
exotic states such as $P_c$ states.

This paper is organized as follows.  
In section \ref{action}, we outline our method of dimensional reduction applied to the Sakai-Sugimoto model with 
extra-dimensional components of the gauge fields identified with heavy mesons. 
In section \ref{cl}, we apply the semi-classical method of collective coordinates to obtain physical baryons. 
Results are compared with experimental data. 

\section{\label{action}Action}

The SU(2+1) (light+heavy flavors) gauge fields on the probed D8 brane in the D4 brane background $ R^{4}\times [0,\infty)\times S^{4} $ have nine components, $ A_{M}(M=0{\rm -}3,U,\alpha) $, 
where $ U $ is a radial coordinate of $ [0,\infty) $ and $ \alpha=\psi,\varphi,\theta_{1},\theta_{2} $ angular coordinates of $ S^{4} $. 
The gauge fields have also flavor components denoted by the index $ a $, where $ A_{M}=A^{a}_{M}\lambda^{a}/2 $ and $ \lambda^{a} $ are the Gell-mann matrices. 
In references~\cite{Sakai2005,2Sakai2005}, the gauge field components on  $ S^{4} $, $ A_{\alpha} $, were ignored. 
In the present work, by regarding $ A^{4{\rm -}7}_{\alpha} $ among $ A^{a}_{\alpha} $ as heavy mesons, we try to introduce heavy flavors in the Sakai-Sugimoto (SS) model. 
When one reduces the dimensions of $ S^{4} $, we will see that the vector fields $ A^{4{\rm -}7}_{\alpha} $ are transformed into scalar heavy meson fields. 
We should remark that 
in embedding the light and heavy mesons in the flavor SU(3) parametrization it is assumed that the SU(3) symmetry is not badly broken for the interaction among them and only their mass difference is the major source of the SU(3) breaking. 
The applicability of such an approach was discussed carefully in Ref.~\cite{Callan1985} in the bound state approach for hyperons. 
Our approach here is also based on the same argument.

Since there are two types of terms in our action, this section is divided into two subsections. 
First, in subsection \ref{FM} we explain how to reduce the dimensions of the higher dimensional Yang-Mills gauge theory. 
Second, in subsection \ref{cs}, we discuss the Chern-Simons term that we will need. 

\subsection{\label{FM}The Yang-Mills part}

In our model, we treat $ A_{\psi}$ and $ A_{\varphi} $ components as heavy meson fields, and try to discuss a system of heavy mesons and nucleons. 
$ A_{\theta_{1}} $ and $ A_{\theta_{2}} $ components are ignored for the minimal use of the extra-dimensional degrees of freedom. 
The nucleons are described by instantons of $ A_{1{\rm -}3,z} $ components~\cite{Witten1998,Douglas1995,Hata2007}. 
To reduce the nine-dimensional theory into a five-dimensional theory, we employ the method proposed by Forg\'{a}cs and Manton~\cite{Manton1980}. 

The YM action that we use is given by the leading term of the DBI action of the SS model~\cite{Sakai2005}, 
\begin{eqnarray}
S^{DBI}_{D8}\simeq T_{8}\left(2\pi^{2}\alpha^{\prime}\right)^{2}\int d^{9}x&&e^{-\phi}\sqrt{-{\rm det}g}g^{MN}g^{PQ} \notag \\
&&\times{\rm tr}\left(\frac{1}{4}F^{{\rm SU(3)}}_{MP}F^{{\rm SU(3)}}_{NQ}\right), \label{action1}
\end{eqnarray}
where $ T_{8}=\left(2\pi\right)^{-8}l^{-9}_{s} $ is the tension of the D8-brane, $ \alpha^{\prime}=l_{s}^{2} $, $ l_{s} $ the string length, and $ \phi $ the dilaton field. 
$ F^{{\rm SU(3)}}_{MN} $ is the field strength of flavor SU(3) gauge fields, and $ F^{{\rm SU(3)}}_{MN}=\partial_{M}A_{N}-\partial_{N}A_{M}+i\left[A_{M},A_{N}\right] $. 
Using the Minkowski metric, $ {\rm diag}\left(\eta^{\mu\nu}\right)=\left(-1,+1,+1,+1\right) $, the metric of the D4 black brane is represented by the follwing $ 9\times9 $ matrix: 
\begin{eqnarray}
&&g^{MN}= \notag \\
&&\left(
    \begin{array}{ccc}
      \left(\frac{R}{U}\right)^{3/2}\eta^{\mu\nu} & 0 & 0 \\
      0 & \left(\frac{U}{R}\right)^{3/2}f & 0 \\
      0 & 0 & \left(\frac{U}{R}\right)^{3/2}U^{-2}g^{\alpha\beta}\left(\Omega_{4}\right) \\
    \end{array}
  \right),
\end{eqnarray}
where $ U $ is the radial coordinate of $ S^{5}\equiv [0,\infty)\times S^{4} $, $ R $ and  $ U_{KK} $ characterize the structure of $ S^{5} $ and $ f=1-U^{3}_{KK}/U^{3} $. 

Now, we assume that the gauge field has an SO(3) spherical symmetry in the $ \left(U, \psi, \varphi\right) $ space. 
It means 
that under SO(3) rotational transformations 
a variation of the gauge (vector) field is absorbed into a gauge transformation. 
To see this point let us first introduce the generators of 
SO(3) transformations $ \xi_{m} $: 
\begin{eqnarray}
&&\xi_{1}=\xi^{M}_{1}\partial_{M}={\rm cos}\varphi\frac{\partial}{\partial \psi}-{\rm cot}\psi \ {\rm sin}\varphi\frac{\partial}{\partial \varphi}, \notag \\
&&\xi_{2}=\xi^{M}_{2}\partial_{M}=-{\rm sin}\varphi\frac{\partial}{\partial \psi}-{\rm cot}\psi \ {\rm cos}\varphi\frac{\partial}{\partial \varphi}, \notag \\
&&\xi_{3}=\xi^{M}_{3}\partial_{M}=\frac{\partial}{\partial \varphi}.
\end{eqnarray}
When one represents the $ (U,\psi,\varphi) $ space as Cartesian coordinate, $ \xi_{m} $ are generators that causes rotations around each axis. 
Applying them to the gauge (vector) field, 
an infinitesimal variation is computed by 
\begin{eqnarray}
 \delta_{\xi_{m}}A_{M}= \epsilon \mathcal{L}_ {\xi_{m}} A_{M}
 \label{eq_SO3_transform}
\end{eqnarray}
where $ \mathcal{L}_ {\xi_{m}}$ denotes the Lie derivative associated with $\xi_{m}$, 
and $\epsilon$ is an infinitesimal parameter. 

Naively, we consider a symmetric field configuration satisfies $ \delta_{\xi_{m}} A_{M}=0 $. 
However, in the case of gauge fields, it is possible to consider a weaker condition, 
by relating the variation $ \delta_{\xi_{m}} A_{M} $ with a gauge transformation, 
$ g={\rm e}^{i \epsilon W_{m}} $ with $ W_{m} $ being an su(3) Lie algebra corresponding to $ \xi_{m} $, 
\begin{eqnarray}
\delta_{W_{m}}A_{M}=\epsilon D_{M}W_{m} 
\label{eq_SU2gauge_transform}
\end{eqnarray}
where $D_{M} \equiv \partial_M  + i[ A_M, \ ]$ is a covariant derivative. 
If an SO(3) infinitesimal transformation (\ref{eq_SO3_transform}) equals an infinitesimal gauge transformation i.e. 
\begin{equation} 
\delta_{\xi_{m}}A_{M}=\delta_{W_{m}}A_{M} , 
\label{eq_symmetry_relation}
\end{equation}
we can make the right hand side zero by a gauge transformation. 
Explicitly, this equation can be written as 
\begin{eqnarray}
\left(\partial_{M}\xi^{N}_{m}\right)A_{N}+\xi^{N}_{m}\partial_{N}A_{M}=\partial_{M}W_{m}+i\left[A_{M},W_{m}\right]. \label{symeq1}
\end{eqnarray}
It is important that for the dimensional reduction, 
the SO(3) space-time symmetry is related to the flavor SU(3) gauge symmetry through (\ref{symeq1})~\cite{Manton1980}. 
This is a symmetry analogous to the hedgehog symmetry for skyrmions and instantons.

In order to perform the dimensional reduction for our purpose, we employ a set of ansatze for field configurations~\cite{Manton1980,Manton1979}, 
\begin{eqnarray}
&&W_{m} =\left(\Phi_{3}\frac{{\rm sin}\varphi}{{\rm sin}\psi}, \ \Phi_{3}\frac{{\rm cos}\varphi}{{\rm sin}\psi}, \ 0\right), \label{Wm} \\
&&A_{\mu,U} =A_{\mu,U}\left(x^{\nu},U\right), \notag \\
&&A_{\psi} = -\Phi_{1}\left(x^{\mu},U\right), \notag \\
&&A_{\varphi} = \Phi_{2}\left(x^{\mu},U\right){\rm sin}\psi-\Phi_{3}{\rm cos}\psi, \label{symgaugefield}
\end{eqnarray}
where $ \Phi_{1,2} $ are a function of $ (x^{\mu},U) $ and $ \Phi_{3} $ a constant. 
Then the constraint (\ref{symeq1}) becomes 
\begin{eqnarray}
&&\left[\Phi_{3},\Phi_{1}\right]=-i\Phi_{2}, \notag \\
&&\left[\Phi_{3},\Phi_{2}\right]=i\Phi_{1}, \notag \\
&&\left[\Phi_{3},A_{\mu,U}\right]=0. \label{symeq2}
\end{eqnarray}

If we substitute (\ref{Wm}) and (\ref{symgaugefield}) for (\ref{action1}), we can perform the integration of 
the higher dimensional manifold $ S^{4} $, resulting in a five-dimensional action. 
The result for the YM part (\ref{action1}) becomes~\cite{Manton1980,Manton1979}: 
\begin{eqnarray}
S_{YM}=\kappa\int d^{4}xdz{\rm tr}&&\left[-\frac{1}{2}K^{-1/3}F^{2}_{\mu\nu}-KF^{2}_{\mu z}\right. \notag \\
&&-\frac{4}{9}\left(D_{\mu}\Phi_{m}\right)^{2}-\frac{4}{9}K^{4/3}\left(D_{z}\Phi_{m}\right)^{2} \notag \\
&&\left.-\frac{16}{81}K^{1/3}\left(i\epsilon_{rst}\Phi_{t}+\left[\Phi_{r},\Phi_{s}\right]\right)^{2}\right], \notag \\ 
\label{action2}
\end{eqnarray}
where $ \kappa=N_{c}\lambda/216\pi^{3}=aN_{c}\lambda $, $ N_{c} $ is a color number, and $ \lambda $ the t'Hooft coupling constant. 
We use the change of variables between  $ U $ and $ z $ by $ U^{3}/U^{3}_{KK}=1+z^{2}=K $. $ R $ and $ U_{KK} $ are expressed by Kaluza-Klein mass $ M_{KK} $~\cite{2Sakai2005}. 
In the following we set $ M_{KK}=1 $. 
We can recover an $ M_{KK} $ dependence by dimensional analysis when needed. 

Now, let us provide a solution for the symmetry relations (\ref{symeq1}) or (\ref{symeq2}). 
A derivation is given in Appendix \ref{ApA} in some detail. 
The results are, 
\begin{eqnarray}
A_{\mu,z}=A^{1}_{\mu,z}\frac{\lambda_{1}}{2}+A^{2}_{\mu,z}\frac{\lambda_{2}}{2}+A^{3}_{\mu,z}\frac{\lambda_{3}}{2}+A^{8}_{\mu,z}\frac{\lambda_{8}}{2}, \label{gaugefield}
\end{eqnarray}
\begin{eqnarray}
\Phi=\frac{1}{2}\left(
    \begin{array}{ccc}
      0 & 0 & \phi_{1} \\
      0 & 0 & \phi_{2} \\
      0 & 0 & 0 \\
    \end{array}
  \right), \ 
\tilde{\Phi}=\frac{1}{2}\left(
    \begin{array}{ccc}
      0 & 0 & 0 \\
      0 & 0 & 0 \\
      \phi^{*}_{1} & \phi^{*}_{2} & 0 \\
    \end{array}
  \right),
  \label{Phifield}
\end{eqnarray}
where $ \Phi=\Phi_{1}+i\Phi_{2}, \ \tilde{\Phi}=\Phi_{1}-i\Phi_{2} $ and $ \phi_{1,2} $ are complex scaler fields~\cite{Manton1979}. 
These expressions imply that the gauge fields $A_{\mu, z}$ correspond to light mesons 
and $\phi_{i}$ to heavy mesons.

Substituting the light-heavy decomposed fields (\ref{gaugefield}) and (\ref{Phifield}) for the Yang-Mills action 
(\ref{action2}), we find 
\begin{eqnarray}
S_{YM}=\kappa\int &&d^{4}xdz
\left\{
{\rm tr}\left[-\frac{1}{2}K^{-1/3}F^{2}_{\mu\nu}-KF^{2}_{\mu z}\right] 
\right. \notag \\
&&
-\frac{4}{9}\left(D_{\mu}\phi\right)^{\dagger}\left(D_{\mu}\phi\right)-\frac{4}{9}K^{4/3}\left(D_{z}\phi\right)^{\dagger}\left(D_{z}\phi\right) \notag \\
&&
\left.
-\frac{16}{81}K^{1/3}\left(\frac{12}{9}-2\phi^{\dagger}\phi+\left(\phi^{\dagger}\phi\right)^{2}\right)^{2}
\right\}. \label{action3}
\end{eqnarray}
where $ \phi^{\dagger}=\left(\phi^{*}_{1},\phi^{*}_{2}\right) $ is a two component SU(2) spinor~\cite{Manton1979}. 
$ F_{\mu\nu,z} $ is the field strength of the SU(2)$ \times $ U(1) gauge fields (\ref{gaugefield}), and $ D_{\mu,z} $ a covariant derivative. 

\subsection{\label{cs} The Chern-Simon part}

It is known that the Wess-Zumino-Witten (WZW) term plays a characteristic role 
for heavy baryon dynamics, providing an attraction (repulsion) between a (anti) heavy meson 
and soliton background~\cite{Callan1985}. 
To introduce the relevant term, we follow the argument of Ref.~\cite{Hata2008} and start from the following 
Chern-Simon (CS) term, 
\begin{eqnarray}
S_{CS}=&&\frac{N_{c}}{24\pi^{2}}\int {\rm tr}\mathcal{F}^{3} \label{CS} \notag \\
{\rm tr}\mathcal{F}^{3}=&&{\rm d}\omega_{5}\left(\mathcal{A}\right) \notag \\
=&&{\rm d}\left[{\rm tr}\left(\mathcal{A}\mathcal{F}^{2}-\frac{i}{2}\mathcal{A}^{3}\mathcal{F}-\frac{1}{10}\mathcal{A}^{5}\right)\right],
\end{eqnarray}
where $ \mathcal{F} $ is the field strength of $ \mathcal{A} $, and the 1-form $ \mathcal{A} $ is $ \mathcal{A}=A_{M}dx^{M}+\hat{A}_{M}dx^{M}(M=0, 1, 2 , 3,z,s) $. 
The variable $ s $ parametrizes a new dimension to define the CS term. 
We have included the U(1) gauge field $\hat{A}_{M} $ that did not appear in (\ref{action3}) because 
it corresponds to $ \omega $ meson and plays an important role in stabilizing the instanton solution~\cite{Hata2008}.

The U(1) term of (\ref{CS}) decomposes 
\begin{eqnarray}
S_{CS}=&&\frac{N_{c}}{24\pi^{2}}\int {\rm tr}F^{3} \notag \\
&&+\frac{N_{c}}{24\pi^{2}}\frac{1}{\sqrt{2N_{f}}}\int\left[3\hat{A}{\rm tr}F^{2}+\frac{1}{2}\hat{A}\hat{F}^{2}\right],
\end{eqnarray}
where in the second term 
we have used the Stokes's theorem to reduce the six-dimensional integral to the five-dimensional one. 
If we choose $ A_{z}=0 $ gauge, omit massive modes, and integrate over $ z $, the first term is 
\begin{eqnarray} \label{WZW1}
\frac{N_{c}}{24\pi^{2}}\int {\rm tr}F^{3}\simeq-\frac{iN_{c}}{240\pi^{2}}\int {\rm tr}\left(UdU^{-1}\right)^{5},
\end{eqnarray}
which is nothing but the WZW term~\cite{Hata2008}. 

If we use a hedgehog solution for baryons in Ref.~\cite{Hata2007,Hata2008}, 
the chiral field $ U $ has the following form: 
\begin{eqnarray}
U|_{s=0}= {\rm exp}\left(
    \begin{array}{cc}
      iH\left(\mathbf{x}\right)\hat{\mathbf{x}}\cdot\boldsymbol{\tau}/f_{\pi} & 0 \\
      0 & 0 \\
    \end{array}
  \right), \label{chiral}
\end{eqnarray}
where $ \hat{\mathbf{x}} $ is a unit vector, 
$ \boldsymbol{\tau} $ a Pauli matrix and $ f_{\pi} $ the decay constant of the pion. 
The choice of $s=0 $ corresponds to the boundary of the six-dimensional manifold on which the WZW term is defined. 
Using the instanton solution, the function $ H\left(\mathbf{x}\right) $ is given as 
\begin{eqnarray}
\int^{+\infty}_{-\infty}dz^{\prime}A^{cl}_{z}\left(\mathbf{x}, z^{\prime}\right)=H\left(\mathbf{x}\right)\hat{\mathbf{x}}\cdot\boldsymbol{\tau}. \label{hedgehog}
\end{eqnarray}
Since the WZW term is identically zero for flavor SU(2), this term (\ref{WZW1}) vanishes without heavy mesons corresponding to  $ \lambda_{4{\rm -}7} $. 
Keeping these components 
$ \varphi\left(\mathbf{x}\right) $ we write as the chiral field: 
\begin{eqnarray}
U|_{s=0}= {\rm exp}\left(
    \begin{array}{cc}
      iH\left(\mathbf{x}\right)\hat{\mathbf{x}}\cdot\boldsymbol{\tau}/f_{\pi} & \varphi\left(\mathbf{x}\right)/f_{H} \\
      \varphi^{\dagger}\left(\mathbf{x}\right)/f_{H} & 0 \\
    \end{array}
  \right), \label{assumechiral}
\end{eqnarray}
where $ f_{H} $ is the decay constant of heavy mesons. 
As we will discuss later, 
the function $ \varphi\left(\mathbf{x}\right) $ corresponds to the lowest eigenmode 
of the heavy meson fields $ \phi $ when expanded in the fifth $z$-dimension.

Substituting (\ref{assumechiral}) for (\ref{WZW1}) we find 
\begin{eqnarray}
&&-i\frac{N_{c}}{240\pi^{2}}\int{\rm tr}\left(U{\rm d}U^{-1}\right)^{5} \notag \\
&&=\frac{iN_{c}}{f^{2}_{H}}\int d^{4}xB^{\mu}\left(\varphi^{\dagger}D_{\mu}\varphi-\left(D_{\mu}\varphi\right)^{\dagger}\varphi\right), \label{CK}
\end{eqnarray}
where $ B_{\mu} $ is the baryon number current by the soliton, 
\begin{eqnarray}
B^{\mu}&&=\frac{\epsilon^{\mu\nu\alpha\beta}}{24\pi^{2}}{\rm tr}\left[\left(U_{\pi}\partial_{\nu}U_{\pi}^{-1}\right)\left(U_{\pi}\partial_{\alpha}U_{\pi}^{-1}\right)\left(U_{\pi}\partial_{\beta}U_{\pi}^{-1}\right)\right], \notag \\
\end{eqnarray}
with $ U_{\pi}={\rm exp}\left(iH\left(\mathbf{x}\right)\hat{\mathbf{x}}\cdot\boldsymbol{\tau}/f_{\pi}\right) $~\cite{Callan1988}. 

\subsection{\label{our_action} The model action}

To complete our program, we need to introduce a mass term in the action, which is 
not easily done in the holographic method of Sakai-Sugimoto. 
Supplementing a mass term our model action is 
\begin{eqnarray}
S=S_{YM}+S_{CS} -m^{2}K^{1/3}\phi^{\dagger}\phi 
\label{YMCS}
\end{eqnarray}
where the function $ K^{1/3} $ is introduced in accordance with (\ref{action3}) in consideration of the curved nature of 
the fifth-dimension.

\section{\label{cl}Classical solutions}

\subsection{\label{instantonsolution}The instanton solutions}

To discuss baryon properties, we follow the semiclassical method, 
that is first we find a time-independent classical solution and then 
quantize it by introducing slowly moving time-dependent variables. 
Because the direction $ z $ is curved and the time component is coupled, 
in Ref.~\cite{Hata2007} 
they performed the $ 1/\lambda $ expansion and obtained a solution for the gauge configuration 
order by order. 
In the leading order, 
the SU(2) $\in$ SU(3) part of the gauge field 
$A^{cl}_{M}(\mathbf{x}, z)$ 
and the U(1) part $\hat A^{cl}_{M}(\mathbf{x}, z)$ 
are obtained as, 
\begin{eqnarray}
&&A^{cl}_{M}\left(\mathbf{x},z\right)=-if\left(\xi\right)g\partial_{M}g^{-1} \label{A^cl_M}, \\
&&g\left(\mathbf{x},z\right)=\frac{\left(z-Z\right)-i\left(\mathbf{x}-\mathbf{X}\right)\cdot\boldsymbol{\tau}}{\xi}, \notag \\
&&\hat{A}^{cl}_{M}=0,
\end{eqnarray}
where $ M=1,2,3,z $, and 
\begin{eqnarray}
&&f\left(\xi\right)=\xi^{2}/\left(\xi^{2}+\rho^{2}\right), \notag \\
&&\xi=\sqrt{\left(\mathbf{x}-\mathbf{X}\right)^{2}+\left(z-Z\right)^{2}}. \notag
\end{eqnarray}
Here the parameters $ \left(\mathbf{X},Z\right) $ and $ \rho $ are the collective coordinates for 
the position (center) and  size of the instanton, respectively. 
In the next to leading order, the time-components of the SU(2) and U(1) gauge field are obtained as,  
\begin{eqnarray}
&&A^{cl}_{0}=0, \\
&&\hat{A}^{cl}_{0}=\frac{1}{8\pi^{2}a}\frac{1}{\xi^{2}}\left[1-\frac{\rho^{4}}{\left(\xi^{2}+\rho^{2}\right)^{2}}\right]. 
\end{eqnarray}

\subsection{\label{modeex}The solution of $\phi$}

In the present model, we have a heavy meson field $\phi(\mathbf{x} , z)$ that also 
posses a time-independent classical solution. 
To find it, 
we first employ a mode expansion~\cite{Sakai2005} by a complete set $ \{\psi_{n}\left(z\right)\} $, 
\begin{eqnarray}
\phi\left(\mathbf{x},z\right)=\sum_{n=0}\varphi_{n}\left(\mathbf{x}\right)\psi_{n}\left(z\right),
\end{eqnarray}
where $ \varphi_{n} $ are two component SU(2) spinors. 
We can choose an arbitrary complete set $ \{\psi_{n}\left(z\right)\} $, and therefore, 
we choose the one to diagonalize the kinetic and mass terms in the four-dimensional space-time. 
Such a complete set $ \{\psi_{n}\left(z\right)\} $ satisfies the following eigenvalue equation: 
\begin{eqnarray}
-\partial_{z}\left(K^{4/3}\partial_{z}\psi_{n}\left(z\right)\right)+m^{2}K^{1/3}\psi_{n}\left(z\right)=\lambda_{n}\psi_{n}\left(z\right). \notag \\ \label{psieveq}
\end{eqnarray}
These eigenstates $\psi_{n}\left(z\right)$ correspond to various meson resonances with their eigenvalues regarded as their squared masses. 
If we consider only the lowest eigenmode, 
the quadratic terms in $ \phi $ of (\ref{action3}) become 
\begin{eqnarray}
\kappa\int d^{4}x\left[-\partial_{\mu}\varphi^{\dagger}\left(\mathbf{x}\right)\partial^{\mu}\varphi\left(\mathbf{x}\right)-m^{2}_{H}\varphi^{\dagger}\left(\mathbf{x}\right)\varphi\left(\mathbf{x}\right)\right], \label{phiaction2}
\end{eqnarray}
where $ M=0,1,2,3,z $, $ m_{H}=\sqrt{\lambda_{0}} $, and we redefine $ \psi=\psi_{0} $, $ \varphi=2/3\varphi_{0} $. 
The mass parameter $ m $ is determined such that $ m_{H} $ becomes the heavy meson mass $ \left(D(1870),B(5279)\right) $.

To proceed further, we apply again the $1/\lambda$ expansion. 
For this purpose, first we rescale the fields as follows~\cite{Hata2007}: 
\begin{eqnarray}
&&\tilde{x}^{0}=x^{0}, \ \tilde{x}^{M}=\lambda^{1/2}x^{M}, \notag \\
&&\tilde{\mathcal{A}}_{0}=\mathcal{A}_{0}, \ \tilde{\mathcal{A}}_{M}=\lambda^{-1/2}\mathcal{A}_{M}, \ \tilde{\varphi}=\lambda^{-1/2}\varphi,
\end{eqnarray}
where $ M=1,2,3,z $. In the following calculations, we omit the tilde for simplicity. 
Then, the action for $ \varphi $, $ S_{\varphi} $ becomes to the leading order of $1/\lambda$ expansion 
\begin{eqnarray}
S_{\varphi} \ 
\sim \ aN_{c} \lambda^1 \int d^{4}x
\left(-\partial_{i}\varphi^{\dagger}\partial^{i}\varphi-\varphi^{\dagger}\left(\int dz\psi^{2}\mathcal{A}^{2}_{M}\right)\varphi\right). \notag \\
\label{phiaction3}
\end{eqnarray}
Using the solution (\ref{A^cl_M}), $ \mathcal{A}^{2}_{M} $ is proportional to identity matrix. 
Therefore, to solve the equation of motion for $\varphi\left(\mathbf{x}\right)$ we can decompose the two component SU(2) spinor $\varphi\left(\mathbf{x}\right)$ into $f\left(\mathbf{x}\right)\chi$, where $\chi$ is a two component spinor.
Then, the resulting equation of motion for $ \varphi $ is given as 
\begin{eqnarray} 
\partial^{2}_{r}f+\frac{2}{r}\partial_{r}f-\left(3\int dz\frac{\psi^{2}\left(z^{2}+r^{2}\right)}{\left(z^{2}+r^{2}+\rho^{2}\right)^{2}}\right)f=0. \notag \\ \label{EOMphi1}
\end{eqnarray}

To solve this equation, It is convenient to rescale the variable $ \xi\rightarrow\rho\xi $. 
First, we should discuss the asymptotic behavior. 
At $ r\rightarrow0 $, the third term of (\ref{EOMphi1}) approaches zero, 
so we note that the asymptotic behavior of $f$ is $ f\sim r^{-1} $. 
Next, 
to see this at $ r\rightarrow\infty $, we multiply (\ref{EOMphi1}) by $r^{2}$.
Then, (\ref{EOMphi1}) becomes 
\begin{eqnarray} \label{EOMphi2}
r^{2}\partial^{2}_{r}f+r\partial_{r}f-\left(3\int dz\psi^{2}\frac{\left(z^{2}/r^{2}+1\right)}{\left(z^{2}/r^{2}+1+1/r^{2}\right)^{2}}\right)f=0. \notag \\
\end{eqnarray}
If $ z $ is small, the integrand of the third term of (\ref{EOMphi2}) is 
\begin{eqnarray}
\frac{\left(z^{2}/r^{2}+1\right)}{\left(z^{2}/r^{2}+1+1/r^{2}\right)^{2}}\sim 1. \notag
\end{eqnarray}
Also if $ z $ is large, that term becomes smaller than 1. 
However, in this case $ \psi $ becomes almost zero, so in this region, the third term does not contribute to the equation of motion (\ref{EOMphi2}). 
Therefore, we can set the third term equals $ 3f $. 
After all, $ f|_{r\rightarrow\infty} $ satisfy 
\begin{eqnarray}
r^{2}\partial^{2}_{r}f+r\partial_{r}f-3f=0. \notag
\end{eqnarray}
Therefore, the asymptotic behavior at $ r\rightarrow\infty $ is 
\begin{eqnarray}
f\sim r^{\frac{-1-\sqrt{13}}{2}}. \notag
\end{eqnarray}
We have solved Eq. (\ref{EOMphi1}) numerically satisfying the above asymptotic behaviors. 
This solution will be used when quantizing the classical solution and obtaining 
the mass formula for physical baryons.

\section{\label{Quantization}Quantization}

In section \ref{cl}, we have solved the static classical solutions of an instanton and $ \phi $. 
In the collective quantization method, we consider the dynamics of a soliton 
in a moduli space parameterized collective coordinates, and 
by regarding them as canonical variables. 

\subsection{\label{collective}Collective coordinates}

In our model, collective coordinates are as follows~\cite{Hata2007}: 
\begin{itemize}
\item Position of the instanton $ \left(\mathbf{X},Z\right) $
\item Size of the instanton $ \rho $
\item SU(2) orientation $ V $
\item Two component SU(2) spinor $ \chi $
\end{itemize}
where $ \left(\mathbf{X},Z\right) $ and $ \rho $ are the position and size of the instanton, respectively, and 
$V$ the SU(2) matrix  corresponding to soliton rotations. 
In addition to these coordinates, we need the two component SU(2) spinor $ \chi $ corresponding 
to the vibration of heavy mesons: 
\begin{eqnarray}
\phi=f\left(\mathbf{x}\right)\psi\left(z\right)\left(
    \begin{array}{c}
      \chi_{1} \\
      \chi_{2} \\
    \end{array}
  \right)=f\left(\mathbf{x}\right)\psi\left(z\right)\chi.
\end{eqnarray} 

These collective coordinates describe time-dependent collective motions of the classical solutions. 
Since the present theory is based on a gauge theory, we need to be a bit careful~\cite{Manton2004}. 
The collective coordinates introduce motions along the gauge orbits that cannot be physical motions. 
These unphysical motions can be removed by the following prescription~\cite{Hata2007}, 
\begin{eqnarray}
&&A_{M}\left(t,x^{N}\right)=VA^{cl}_{M}\left(x^{N};X^{N}\left(t\right),\rho\left(t\right)\right)V^{-1}-iV\partial_{M}V^{-1}, \notag \\
\label{AMwithcollmotion}\\
&&\phi\left(t,x^{N}\right)=V\phi^{cl}\left(x^{N};\rho\left(t\right),\chi\left(t\right)\right), \label{phisolution}
\end{eqnarray}
where $ V=V\left(t, x^{N}\right) $ is an element of the gauge group SU(2). 
In the $ A_{0}=0 $ gauge with imposing the Gauss's law: 
\begin{eqnarray}
D^{cl}_{M}\left(\dot{X}^{N}\frac{\partial}{\partial X^{N}}A^{cl}_{M}+\dot{\rho}\frac{\partial}{\partial\rho}A^{cl}_{M}-D^{cl}_{M}\Phi\right)=0,
\label{Gausslaw}
\end{eqnarray}
where $ M,N=1,2,3,z $, $ \Phi=-iV^{-1}\dot{V} $ and $ D^{cl}_{M}=\partial_{M}+i\left[A^{cl}_{M}, \ \right] $. 
By having the solution of $\Phi$ to (\ref{Gausslaw})~\cite{Hata2007}, spurious motions along the gauge orbits 
are removed in the collective motions of (\ref{AMwithcollmotion}).

\subsection{\label{heavymass}Heavy meson field}

The action for the heavy meson field $ \varphi $, $ S_{\varphi} $, is 
\begin{eqnarray}
S_{\varphi}=&&aN_{c}\int d^{4}x\left[\lambda^{1}\left(-\partial_{i}\varphi^{\dagger}\partial^{i}\varphi-\varphi^{\dagger}\left(\int dz\psi^{2}\mathcal{A}^{2}_{M}\right)\varphi\right)\right. \notag \\
&&\left.+\lambda^{0}\left(\int dz\psi^{2}\left(D_{0}\varphi\right)^{\dagger}D_{0}\varphi-m^{2}_{H}\varphi^{\dagger}\varphi\right)\right] \notag \\
&&+\lambda^{0}\frac{iN_{c}}{f^{2}_{H}}\int d^{4}xB^{\mu}\left(\varphi^{\dagger}D_{\mu}\varphi-\left(D_{\mu}\varphi\right)^{\dagger}\varphi\right), \label{phiaction4}
\end{eqnarray}
where the covariant derivative $ D_{0} $ is defined as $ D_{0}\varphi=\partial_{0}\varphi+i\mathcal{A}_{0}\varphi $.

It is convenient to introduce the heavy meson field as~\cite{Manohar2000} 
\begin{eqnarray}
\phi=e^{\mp im_{H}t}\tilde{\phi}=f\left(\mathbf{x}\right)\psi\left(z\right)e^{\mp im_{H}t}\tilde{\chi}\left(t\right),
\end{eqnarray}
where $ -/+ $ correspond to heavy/anti-heavy mesons. 
Then, if we only consider the leading terms of $ 1/m_{H} $ expansion and substitute the solutions (\ref{phisolution}) for (\ref{phiaction4}), the first line of (\ref{phiaction4}) is zero and the second line becomes 
\begin{eqnarray}
&&\int d^{4}xdzf^{2}\psi^{2}\left[\left(D_{0}\left(V\chi\right)\right)^{\dagger}D_{0}\left(V\chi\right)-m^{2}_{H}\chi^{\dagger}\chi\right] \notag \\
&&\simeq2m_{H}\int d^{4}xdzf^{2}\psi^{2}\tilde{\chi}^{\dagger}D_{0}\tilde{\chi}.
\end{eqnarray}

\subsection{\label{quantization}Quantization}

By employing the normalization $ aN_{c}\int d^{3}xdz f^{2}\psi^{2}=1 $, absorbing the coefficient of the kinetic term of $ \tilde{\chi} $ and integrating over the space of $ (x^{\mu},z) $, 
finally we obtain the action of collective motions as follows~\cite{Hata2007}: 
\begin{eqnarray}
&&\int dt\left[L_{X}+L_{Z}+L_{y}\right]+\mathcal{O}\left(\lambda^{-1},m_{H}^{-1}\right), \notag \\
&&L_{X}=-M_{0}+\frac{m_{X}}{2}\dot{\mathbf{X}}^{2}, \notag \\
&&L_{Z}=\frac{M_{Z}}{2}\dot{Z}^{2}-\frac{m_{Z}\omega^{2}_{Z}}{2}Z^{2}, \notag \\
&&L_{y}=\frac{m_{y}}{2}\dot{y}^{2}_{I}-\frac{m_{y}\omega^{2}_{\rho}}{2}\rho^{2}-\frac{Q}{\rho^{2}}, \notag \\
&&L_{\chi}=\pm i\tilde{\chi}^{\dagger}\partial_{t}\tilde{\chi}\pm A\frac{N_{c}}{\rho^{2}}\tilde{\chi}^{\dagger}\tilde{\chi}, \label{lagrangian}
\end{eqnarray}
with 
\begin{eqnarray}
&&M_{0}=8\pi^{2}\kappa, \notag \\
&&m_{X}=m_{Z}=8\pi^{2}aN_{c}, \ m_{y}=16\pi^{2}aN_{c}, \notag \\
&&\omega^{2}_{Z}=\frac{2}{3}, \ \omega^{2}_{\rho}=\frac{1}{6}, \ Q=\frac{N^{2}_{c}}{40\pi^{2}a}. \label{quantumcoefficient}
\end{eqnarray}
Also if we consider only the leading terms of $ 1/m_{H} $ expansion, then only the time component contributes and therefore $ B^{\mu} $ becomes 
\begin{eqnarray}
B^{0}=\frac{1}{2\pi^{2}}\frac{{\rm sin}^{2}H}{r^{2}}\frac{dH}{dr}.
\end{eqnarray}
Therefore, $ A $ is written as 
\begin{eqnarray}
A=\frac{4}{\pi f^{2}_{H}}\int dr{\rm sin}^{2}H\frac{dH}{dr}f^{2}-a\int d^{3}xdz\hat{A}^{cl}_{0}\psi^{2}f^{2}, \notag \\ \label{A}
\end{eqnarray}
where using the rescale $ x^{M}\rightarrow\rho x^{M} $, $ \hat{A}^{cl}_{0} $ is 
\begin{eqnarray}
\hat{A}^{cl}_{0}=\frac{1}{8\pi^{2}a}\frac{1}{\xi^{2}}\left[1-\frac{1}{\left(\xi^{2}+1\right)^{2}}\right].
\end{eqnarray}
Therefore, we quantize the system with the canonical variables of these collective coordinates.

Before deriving the mass formula, we consider quantum numbers of the heavy mesons~\cite{Callan1985}. 
In the classical solution, the heavy meson fields have the spin 0 and the isospin 1/2. 
First, we consider the isospin rotation. 
By the isospin rotation $ g_{I}=e^{i\boldsymbol{\theta}\cdot\mathbf{I}} $, the gauge fields are transformed into 
\begin{eqnarray}
A_{M}&&\rightarrow g_{I}A_{M}g^{-1}_{I}-ig_{I}\partial_{M}g^{-1}_{I} \notag \\
&&=\left(g_{I}V\right)A^{cl}_{M}\left(g_{I}V\right)^{-1}-i\left(g_{I}V\right)\partial_{M}\left(g_{I}V\right)^{-1}.
\end{eqnarray}
On the other hand, the heavy meson field transforms 
\begin{eqnarray}
V\tilde{\chi}\rightarrow g_{I}V\tilde{\chi}.
\end{eqnarray}
Therefore, $ V $ carries the isospin, and $ V $ and $ \tilde{\chi} $ have the following transformation properties: 
\begin{eqnarray}
\left\{
    \begin{array}{c}
      V\rightarrow g_{I}V \\
      \tilde{\chi}\rightarrow\tilde{\chi}. \\
    \end{array}
\right.
\end{eqnarray}

Second, we consider the spatial rotation. 
When the gauge transformation which is equivalent to the spatial rotation is written as $ g_{J}=e^{i\boldsymbol{\theta}\cdot\mathbf{J}} $, spatial rotation act the gauge field as follows: 
\begin{eqnarray}
&&A_{M}\left(t,R_{NP}x^{P}\right) \notag \\
&&=VA^{cl}_{M}\left(R_{NP}x^{P};R_{NP}X^{P}\right)V^{-1}-iVV^{-1} \notag \\
&&=\left(Ve^{-i\boldsymbol{\theta}\cdot\mathbf{I}}\right)A^{cl}_{M}\left(x^{N};X^{N}\right)\left(Ve^{i\boldsymbol{\theta}\cdot\mathbf{I}}\right)^{-1} \notag \\
&& \ \ \ \ \ \ \ \ \ \ \ \ \ \ -i\left(Ve^{-i\boldsymbol{\theta}\cdot\mathbf{I}}\right)\partial_{M}\left(Ve^{i\boldsymbol{\theta}\cdot\mathbf{I}}\right)^{-1},
\end{eqnarray}
where the hedgehog like structure relates the spatial rotation to isospin rotation, and so the spatial rotation is expressed by $ g_{I} $. 
Also, the scalar field is transformed into 
\begin{eqnarray}
&&V\tilde{\phi}\left(t,x\right)\rightarrow V\tilde{\phi}\left(t,R_{MN}x^{N}\right)=Ve^{i\boldsymbol{\theta}\cdot\mathbf{J}}\tilde{\phi} \notag \\
&&=Ve^{-i\boldsymbol{\theta}\cdot\mathbf{I}}f\psi e^{i\boldsymbol{\theta}\cdot\mathbf{T}}\tilde{\chi}\left(t\right),
\end{eqnarray}
where $ \mathbf{T}=\mathbf{J}+\mathbf{I} $ is the grand spin operator. 
Therefore, $ V $ and $ \chi $ have the following transformation properties: 
\begin{eqnarray}
\left\{
    \begin{array}{c}
      V\rightarrow Ve^{-i\boldsymbol{\theta}\cdot\mathbf{I}} \\
      \tilde{\chi}\rightarrow e^{i\boldsymbol{\theta}\cdot\mathbf{T}}\tilde{\chi}. \\
    \end{array}
\right.
\end{eqnarray}
From the above, after doing the collective rotation, $ \tilde{\chi} $ has the spin 1/2 and the isospin 0. 
Thus, we should quantize $ \tilde{\chi} $ as fermions: 
\begin{eqnarray}
\left\{\tilde{\chi}_{i},\tilde{\chi}^{\dagger}_{j}\right\}=\tilde{\chi}_{i}\tilde{\chi}^{\dagger}_{j}+\tilde{\chi}^{\dagger}_{j}\tilde{\chi}_{i}=\delta_{ij}.
\end{eqnarray}

\subsection{\label{massformula}Mass formula}

By collecting the terms proportional to $1/\rho^2$ in (\ref{lagrangian}), our collective Hamiltonian 
takes essentially the same form as that of Ref.~\cite{Hata2007}. 
Therefore,  we can follow the same quantization procedure, resulting in 
the mass formula as follows, 
\begin{eqnarray}
M=&&M_{0}+\left(N_{Q}+N_{\overline{Q}}\right)m_{H} \notag \\
&&+\sqrt{\frac{\left(l+1\right)^{2}}{6}+\frac{2N^{2}_{c}}{15}\left(1-\frac{40a\pi^{2}A}{N_{c}}\left(N_{Q}-N_{\overline{Q}}\right)\right)}M_{KK} \notag \\
&&+\frac{2\left(n_{\rho}+n_{Z}\right)+2}{\sqrt{6}}M_{KK} , \label{massformula1}
\end{eqnarray}
where $ M_{0} $ is the instanton mass, $ N_{Q/\overline{Q}} $ the number of heavy/anti-heavy mesons. 
We emphasize that our mass formula contains a numerical constant $A$ 
in the second line.  This is a unique feature of our model construction. 
As we will see shortly,  this term plays a crucial role in reproducing 
the mass ordering of $\Sigma_c$ and $\Lambda_c^*$ correctly as in experimental 
data. 

We find that the spin $ \mathbf{J} $ and the isospin $ \mathbf{I} $ of the instanton are both $ l/2 $. 
The spin of the baryons is the sum of spins of the instanton and heavy mesons. 
In the SS model, parity transformation is defined by $ x^{M}\rightarrow-x^{M} $. 
Also, when $ n_{Z} $ is even or odd, the wave function of $ n_{Z} $ has parity even or odd. 
Therefore, the quantum numbers $ \left(n_{\rho},n_{z}\right) $ correspond 
to  radial excitations and those which flip parity, respectively. 
Note that $ \tilde{\phi}=f\psi\tilde{\chi} $ has parity even.

\begin{table}[h]
  \caption{Parameters in our model.}
  \label{parameter}
  \centering
  \begin{tabular}{ccccc}
    \hline
    $M_{0}$(MeV) & $M_{KK}$(MeV) & $m/M_{KK}$ & $m/M_{KK}$ & $f_{\pi}/M_{KK}$ \\
    \hline
    -572 & 500 &4.385 & 10.62 & 0.122 \\
     & & (charm) & (bottom) & \\    
    \hline \label{parameter}
  \end{tabular}
\end{table}

Parameters in or mass are $ \left(M_{0},M_{KK},m,f_{\pi}\right) $ and are  shown in TABLE \ref{parameter}. 
As explained in section \ref{cl}, $ m $ is determined such that $ m_{H} $ becomes the heavy meson mass $ \left(D(1870),B(5279)\right) $. 
Also, we use $ f_{D}/f_{\pi}=1.7 $ and $ f_{B}/f_{\pi}=1.6 $ as in Ref.~\cite{Dominguez1987}, 
and the pion decay constant is $ f_{\pi}\times M_{KK}=61.2 \ {\rm MeV} $ which is about 30\% 
smaller than the experimental value 93.2 MeV~\cite{Dominguez1987}. 
For the Kalza-Klein mass, we use $ M_{KK} =500$ MeV which is the same value as in Ref. \cite{Hata2007}. 
Having these inputs, there is only one free parameter $M_0$ which is fixed to reproduced 
the mass of $ \Lambda_{c}(2286) $. 
We are then interested in mass differences of baryons. 
We note that our mass formula (\ref{massformula1}) differs from that of Ref.~\cite{Liu2017} in the term proportional to $A$. 
In our model, the term depends on the heavy meson decay constant $ f_{H} $, while that of Ref.~\cite{Liu2017} does not have such parameter dependence. 
From (\ref{A}) with the decay constant values as in TABLE \ref{parameter}, we find $ A=0.078 $ for charm and $ A=3.7 $ for bottom sectors, respectively.

Results are summarizing in TABLE \ref{massspectra}.
\begin{widetext}
\begin{table}[h]
  \caption{Charmed and Bottomed baryons}
  \label{massspectra}
  \centering
  \begin{tabular}{ccccccccc}
    \hline
    $ B $ & $ IJ^{P} $ & $ \ \ \ \ \ l \ \ \ \ \ $ & $ \ \ \ \ \ n_{\rho} \ \ \ \ \ $ & $ \ \ \ \ \ n_{z} \ \ \ \ \ $ & $ \ \ \ \ \ N_{Q} \ \ \ \ \ $ & $ \ \ \ \ \ N_{\overline{Q}} \ \ \ \ \ $ & our model/MeV & exp./MeV \\
    \hline \hline
    $ \Lambda_{c} $ & $ 0\frac{1}{2}^{+} $ & 0 & 0 & 0 & 1 & 0 & [2286] & 2286 \\
    $ \Sigma_{c} $ & $1\frac{1}{2}^{+} $ & 2 & 0 & 0 & 1 & 0 & 2523 & $ 2453 $ \\
     & $1\frac{3}{2}^{+} $ & 2 & 0 & 0 & 1 & 0 & 2523 & $ 2520 $ \\   
    $ \Lambda^{*}_{c} $ & $ 0\frac{1}{2}^{-} $ & 0 & 0 & 1 & 1 & 0 & 2694 & (2595) \\    
     & $ 0\frac{1}{2}^{+} $ & 0 & 1 & 0 & 1 & 0 & 2694 &  (2765) \\    
    $ \Sigma^{*}_{c} $ & $ 1\frac{1}{2}^{-},1\frac{3}{2}^{-} $ & 2 & 0 & 1 & 1 & 0 & 2931 & -\\
     & $ 1\frac{1}{2}^{+},1\frac{3}{2}^{+} $ & 2 & 1 & 0 & 1 & 0 & 2931 & -\\
    $ P_{c} $ & $ \frac{1}{2}\frac{1}{2}^{-},\frac{1}{2}\frac{3}{2}^{-} $ & 1 & 0 & 0 & 1 & 1 & 4255 & 4312/4380/4440,4457 \\    
    $ P^{*}_{c} $ & $ \frac{1}{2}\frac{1}{2}^{-},\frac{1}{2}\frac{3}{2}^{-} $ & 1 & 0 & 1 & 1 & 1 & 4664 & - \\    
     & $ \frac{1}{2}\frac{1}{2}^{+},\frac{1}{2}\frac{3}{2}^{+} $ & 1 & 1 & 0 & 1 & 1 & 4663 & - \\    
    $ \Lambda_{b} $ & $ 0\frac{1}{2}^{+} $ & 0 & 0 & 0 & 1 & 0 & 5676 & 5620 \\
    $ \Sigma_{b} $ & $1\frac{1}{2}^{+} $ & 2 & 0 & 0 & 1 & 0 & 5919 & $ 5810 $ \\
     & $1\frac{3}{2}^{+} $ & 2 & 0 & 0 & 1 & 0 & 5919 & $ 5830 $ \\    
    $ \Lambda^{*}_{b} $ & $ 0\frac{1}{2}^{-} $ & 0 & 0 & 1 & 1 & 0 & 6084 & $ 5912 $ \\    
     & $ 0\frac{1}{2}^{+} $ & 0 & 1 & 0 & 1 & 0 & 6084 & (6072) \\    
    $ \Sigma^{*}_{b} $ & $ 1\frac{1}{2}^{-},1\frac{3}{2}^{-} $ & 2 & 0 & 1 & 1 & 0 & 6327 & - \\
     & $ 1\frac{1}{2}^{+},1\frac{3}{2}^{+} $ & 2 & 1 & 0 & 1 & 0 & 6327 & - \\
    $ P_{b} $ & $ \frac{1}{2}\frac{1}{2}^{-},\frac{1}{2}\frac{3}{2}^{-} $ & 1 & 0 & 0 & 1 & 1 & 11070 & - \\    
    $ P^{*}_{b} $ & $ \frac{1}{2}\frac{1}{2}^{-},\frac{1}{2}\frac{3}{2}^{-} $ & 1 & 0 & 1 & 1 & 1 & 11480 & - \\    
     & $ \frac{1}{2}\frac{1}{2}^{+},\frac{1}{2}\frac{3}{2}^{+} $ & 1 & 1 & 0 & 1 & 1 & 11480 & - \\    
    \hline
  \end{tabular}
\end{table}
\end{widetext}

These results have some characteristic properties as follows. 
\begin{itemize}
\item The quantum numbers $ (n_{\rho},n_{z})$ physically correspond to the Roper and the odd parity excitations. 
As observed in the previous works~\cite{Hata2007}, the mass formula (\ref{massformula1}) indicates the degeneracy between them which agrees well 
with experimental data for the light flavor sector,  the feature that is difficult to be explained by a naive quark model. 
This feature seems to be generalized to strange baryons~\cite{Takayama1999}. 
Whether this also extends to charm and bottom sectors is an interesting question. 
Possible candidates are $\Lambda_c(2765)$ and $\Lambda_b(6072)$, while their spin and parity are to be determined. 
\item If we expand the mass formula (\ref{massformula1}) by $ 1/N_{c} $, the mass splitting of $ \Lambda_{c} $ and $ \Sigma_{c} $ is proportional to $ 1/N_{c} $. 
This splitting is related to the spin-spin interaction, and the $N_{c}$ dependency is consistent to that of the $ 1/N_{c} $ expansion scheme. 
\item Because we have included only the leading terms of $ 1/m_{H} $ expansion, 
we have obtained the heavy quark symmetry (HQS) singlet $ \Lambda_{c,b}(0\frac{1}{2}^{+}) $ and the doublet $ \Sigma_{c,b}(1\frac{1}{2}^{+}, 1\frac{3}{2}^{+}) $. 
On the other hand, 
the lowest $\Lambda_{c,b}(0\frac{1}{2}^{-})$ and $\Lambda_{c,b}(0\frac{3}{2}^{-})$ in a quark model do not exist in the present model, 
because this state is considered to correspond to the $ \lambda $ mode. 
This is the reason that we put the mass value 2595 in parentheses in TABLE \ref{massspectra}. 
In our mass formula the excited states of baryons are described by excitations of the instanton, 
which correspond to $ \rho $ modes in a quark model language. 
\item Empirically, the mass splitting of $ \Lambda_{c} $ and $ \Lambda^{*}_{c} $ is about twice larger than that of $ \Lambda_{c} $ and $ \Sigma_{c} $. 
In the present study, the value of $A$ plays an important role to make this order of baryon masses. 
In particular, for 
$\Delta_{\Sigma_c - \Lambda_c} \equiv M(\Sigma_c) - M(\Lambda_c)$ and 
$\Delta_{\Lambda_c^* - \Lambda_c} \equiv M(\Lambda_c^*) - M(\Lambda_c)$, 
we have 
$\Delta_{\Sigma_c - \Lambda_c} < \Delta_{\Lambda_c^* - \Lambda_c}$
in accordance with the experimental data, while the formula in ~\cite{Liu2017} results in the reversed relation. 
Let $B$ be 
\begin{eqnarray}
B=1-\frac{40a\pi^{2}A}{N_{c}}.
\end{eqnarray}
For $ B=0 $, we find $ \Delta_{\Sigma_c - \Lambda_c}=\Delta_{\Lambda_c^* - \Lambda_c} $. As $ A $ becomes smaller (i.e. $B$ becomes larger), 
$ \Delta_{\Sigma_c - \Lambda_c} $ becomes larger, and at some point, $\Delta_{\Lambda_c^* - \Lambda_c} $ equals $ 2\Delta_{\Sigma_c - \Lambda_c}$. 
\item Our model has hidden charmed pentaquark states corresponding to $ P_{c}(4312/4380/4440,4457) $ states reported recently~\cite{Pc2015,Pc}. 
Similarly, we predicted the mass of hidden bottomed pentaquark states not yet observed, which we denote by $ P_{b} $ here. 
Since the soliton has the same value of the spin and the isospin which stems from the hedgehog structure, our mass formula cannot generate the $ P_{c} $ state with the spin $\frac{1}{2}\frac{5}{2}^{+}$. 
\end{itemize}

\section{\label{conclusion}Conclusion}

In this paper, we aimed to apply the holographic model proposed by Sakai and Sugimoto to heavy flavor baryons, and derive a mass formula. 
In our model, extra-dimensional components of the gauge fields omitted in Ref.~\cite{Sakai2005,2Sakai2005} have been interpreted as heavy mesons. 
The gauge fields living in $ S^{4} $ have been transformed into heavy mesons by the Forg\'acs-Manton method, and we have obtained the action composed of light and heavy mesons. 
Then, heavy baryons have represented as composite states of heavy mesons and an instanton composed by light flavors. 
In addition to the collective coordinates used in~\cite{Hata2007}, our model has the dynamical variable corresponding to the vibration of heavy mesons. 
We have performed the collective coordinate quantization of the system consisted of these coordinates, and obtained the mass formula of heavy baryons. 
When quantizing our model, as in Ref.~\cite{Callan1985}, 
heavy mesons behave as heavy quarks, which was referred 
to as transmutation of quantum numbers in the intrinsic 
frame of the hedgehog instanton~\cite{Liu2017}. 
The mass formula has given the mass spectra that are compared with existing experimental data.

In our model, we have considered the limit of the large $ N_{c} $ and the t'Hooft coupling $ \lambda $ as in~\cite{Hata2007}, 
and also took the limit of large $ m_{H} $. 
We have treated the only leading terms of $ 1/m_{H} $, so we have obtained the HQS singlet $ \Lambda_{c,b}(0\frac{1}{2}^{+}) $ and the doublet $ \Sigma_{c,b}(1\frac{1}{2}^{+}, 1\frac{3}{2}^{+}) $.  
Also, our mass formula has yielded the degenerate Roper like and odd parity excitations. 
Moreover, we have realized the mass ordering $\Delta_{\Sigma_c - \Lambda_c} < \Delta_{\Lambda_c^* - \Lambda_c}$ 
in accordance with the experimental data. 
Furthermore, our model has hidden charmed pentaquark states $ P_{c}(4312/4380/4440,4457) $ reported recently~\cite{Pc2015,Pc}. 
Similarly, we have predicted the masses of hidden bottomed pentaquark states not yet observed.

\begin{acknowledgments}
This work is supported in part by JSPS KAKENHI No. JP17K05441 (C) 
and Grants-in-Aid for Scientific Research on Innovative Areas (No. 18H05407). 
\end{acknowledgments}

\appendix

\section{Solving constraints} \label{ApA}

In this Appendix, we will solve following constraints: 
\begin{eqnarray}
&&\left[\Phi_{3},\Phi\right]=-\Phi, \label{Asymeq4} \\
&&\left[\Phi_{3},\tilde{\Phi}\right]=\tilde{\Phi}, \label{Asymeq5} \\
&&\left[\Phi_{3},A_{\mu,z}\right]=0, \label{Asymeq7}
\end{eqnarray}
where $ \Phi=\Phi_{1}+i\Phi_{2} $, $ \tilde{\Phi}=\Phi_{1}-i\Phi_{2}$. 
By using the Gell-mann matrices $ \lambda_{a} $, we can define the Cartan's standard form: 
\begin{eqnarray}
&&H_{1}=\frac{\lambda_{3}}{2}, \ H_{2}=\frac{\lambda_{8}}{2}, \notag \\
&&E_{\pm\boldsymbol{\gamma}}=\frac{1}{2}\left(\frac{\lambda_{1}}{2}\pm i\frac{\lambda_{2}}{2}\right), \notag \\
&&E_{\pm\boldsymbol{\alpha}}=\frac{1}{2}\left(\frac{\lambda_{4}}{2}\pm i\frac{\lambda_{5}}{2}\right), \notag \\
&&E_{\pm\boldsymbol{\beta}}=\frac{1}{2}\left(\frac{\lambda_{6}}{2}\pm i\frac{\lambda_{7}}{2}\right),
\end{eqnarray}
where roots $ \left(\boldsymbol{\alpha},\boldsymbol{\beta},\boldsymbol{\gamma}\right) $ are given as 
\begin{eqnarray}
&&\pm\boldsymbol{\alpha}=\left(\pm\frac{1}{2},\pm\frac{\sqrt{3}}{2}\right), \notag \\
&&\pm\boldsymbol{\beta}=\left(\mp\frac{1}{2},\pm\frac{\sqrt{3}}{2}\right), \notag \\
&&\pm\boldsymbol{\gamma}=\left(\pm1,0\right).
\end{eqnarray}
Furthermore, we define $ h_{\boldsymbol{\omega}} $: 
\begin{eqnarray}
h_{\boldsymbol{\omega}}=\omega^{i}H_{i}.
\end{eqnarray}

In order to solve (\ref{Asymeq7}), if we write $ \Phi_{3}=\Phi^{i}_{3}H_{i} $, then we have 
\begin{eqnarray}
\left[\Phi_{3},E_{\boldsymbol{\omega}}\right]=\Phi^{i}_{3}\omega_{i}E_{\boldsymbol{\omega}}.
\end{eqnarray}
Therefore, when one chooses $ \Phi^{i}_{3} $ appropriately, $ \Phi_{3} $ commutes with $ E_{\boldsymbol{\omega}} $ for a root $ \boldsymbol{\omega} $. 
We choose $ \Phi^{i}_{3} $ as to commute with $ E_{\boldsymbol{\gamma}} $ here. 
Then, all generators commuting with $ \Phi_{3} $ are $ E_{\boldsymbol{\gamma}},E_{-\boldsymbol{\gamma}},h_{\boldsymbol{\gamma}},h $, where $ h $ satisfies $ {\rm tr}\left(h_{\boldsymbol{\gamma}}h\right)=0 $ and $ {\rm tr}\left(hh\right)=1/2 $. 
Therefore, if $ A_{\mu, z} $ are written as the linear combination of those generators, (\ref{Asymeq7}) is satisfied. 
This is the gauge field on the subgroup SU(2)$\times $U(1), we can written as 
\begin{eqnarray}
A_{\mu,z}=A^{1}_{\mu,z}\frac{\lambda_{1}}{2}+A^{2}_{\mu,z}\frac{\lambda_{2}}{2}+A^{3}_{\mu,z}\frac{\lambda_{3}}{2}+A^{8}_{\mu,z}\frac{\lambda_{8}}{2}. \label{Agaugefield}
\end{eqnarray}

Next, to solve (\ref{Asymeq4}) and (\ref{Asymeq5}), we choose $ \Phi^{i}_{3} $ such that we have 
\begin{eqnarray}
\left[\Phi_{3},E_{\boldsymbol{\alpha}}\right]=\Phi^{i}_{3}\alpha_{i}E_{\boldsymbol{\alpha}}=-E_{\boldsymbol{\alpha}}.
\end{eqnarray}
Due to $ \boldsymbol{\beta}=\boldsymbol{\alpha}-\boldsymbol{\gamma} $, if we write the form, 
\begin{eqnarray}
&&\Phi=\phi_{1}E_{\boldsymbol{\alpha}}+\phi_{2}E_{\boldsymbol{\beta}}, \notag \\
&&\tilde{\Phi}=\tilde{\phi}_{1}E_{-\boldsymbol{\alpha}}+\tilde{\phi}_{2}E_{-\boldsymbol{\beta}},
\end{eqnarray}
(\ref{Asymeq4}) and (\ref{Asymeq5}) are satisfied, where since $ \Phi_{1,2} $ is Hermitian matrices, we can show $\tilde{\phi}_{1}=\phi^{*}_{1} $ and $ \tilde{\phi}_{2}=\phi^{*}_{2} $.

\bibliography{apssamp}

\end{document}